\documentclass[aps,prx,twocolumn,groupedaddress, floatfix,longbibliography,nofootinbib,byrevtex]{revtex4}
\usepackage{graphicx}
\usepackage{dcolumn}
\usepackage{bm}
\usepackage[utf8]{inputenc}
\usepackage{amsmath,amsfonts,calc}
\usepackage[pagebackref = true, colorlinks, linkcolor = Green, citecolor = Blue, bookmarksdepth=2, linktocpage=true]{hyperref}
\usepackage[capitalize]{cleveref}
\usepackage{comment}
\usepackage[normalem]{ulem}
\usepackage[usenames,dvipsnames]{color}
\usepackage{graphicx}
\usepackage{scalerel}
\usepackage{mathtools}
\usepackage{stackengine,wasysym}
\makeatletter
\newsavebox\myboxA
\newsavebox\myboxB
\newlength\mylenA
\makeatother

\begin{document}
\title{On the static effective Lindbladian of the squeezed Kerr oscillator}
\author{Jayameenakshi Venkatraman}
\email{jaya.venkat@yale.edu,xu.xiao@yale.edu,\\rodrigo.cortinas@yale.edu}
\thanks{these three authors contributed equally.}
\author{Xu Xiao}
\email{jaya.venkat@yale.edu,xu.xiao@yale.edu,\\rodrigo.cortinas@yale.edu}
\thanks{these three authors contributed equally.}
\author{Rodrigo G. Cortiñas}
\email{jaya.venkat@yale.edu,xu.xiao@yale.edu,\\rodrigo.cortinas@yale.edu}
\thanks{these three authors contributed equally.}
\author{Michel H. Devoret}
\email{michel.devoret@yale.edu}
\affiliation{Departments of Applied Physics and Physics, Yale University, New Haven, CT 06520, USA}
\date{\today}
\begin{abstract}
We derive the static effective Lindbladian beyond the rotating wave approximation (RWA) for a driven nonlinear oscillator coupled to a bath of harmonic oscillators. The associated dissipative effects may explain orders of magnitude differences between the predictions of the ordinary RWA model and results from recent superconducting circuits experiments on the Kerr-cat qubit. The higher-order dissipators found in our calculations have important consequences for quantum error-correction protocols and parametric processses.
\end{abstract}
\maketitle

\subsection*{Introduction}

Static effective Hamiltonians can be engineered in circuit quantum electrodynamics \cite{blais2021} by coherently driving parametric processes. Such technique has been put to use in creating qubits \cite{leghtas2015, wang2016,lescanne2020,puri2017}, gates between them \cite{Kurpiers2018,Axline2018,Rosenblum2018,Gao2018}, readout schemes \cite{krantz2016,eddins2018,touzard2019}, and quantum simulations \cite{kandala2017,boettcher2020,altman2021,wang2020}. Similar techniques are employed in quantum simulation with atomic systems \cite{goldman2014,Wintersperger2020,Martinez2016}. Effective Hamiltonians resulting from complex pulse sequences in Trotterization schemes applied to a system \cite{Campagne-Ibarcq2020,Royer2020,Martinez2016,Eickbusch2021} can be also viewed as belonging to the above class. 

Since physical systems are inevitably open, the nonlinear mixing processes associated with the Hamiltonian parametric terms of interest are also driven incoherently by fluctuations of the environment. These environmental fluctuations can be thermal in origin, in which case the process can be understood as a classical nonlinear mixing of noise that is down- or up-converted to the frequency of the nonlinear oscillator, or have an origin in the vacuum fluctuations of the environment. These vacuum fluctuations can be amplified by the drive and give rise to heating even in a zero temperature environment, a phenomenon known as Unruh heating when the driving force produces a simple time-independent acceleration \cite{Wilson2011,Blencowe2020,Unruh1976}.

A recent work \cite{petrescu2020} studied these effects in an attempt to explain drive-induced lifetime reduction in transmon circuits during readout. But in transmons, these effects tend to be masked by multiphoton nonlinear resonances limiting readout and  parametric operations \cite{sank2016,blais2021,Shillito2022,Cohen2022}. However, the recent implementation of a squeezed Kerr oscillator giving rise to the Kerr-cat qubit \cite{puri2017,grimm2020,frattini2022} provides an ideal platform to uncover the effect of drive-enhanced environmental fluctuations, since unwanted nonlinear resonances of the transmon qubit are largely absent in this new system. Mixing of the environmental fluctuations is captured by beyond rotating wave approximation (RWA) in corrections to the system-bath coupling, giving rise to modified Lindbladian dynamics. In this note, we 
compute the static effective dissipators for the Kerr-cat system and discuss possible new effects that may explain experimental data in \cite{frattini2022}. Our systematic method, based on \cite{Venkatraman2021}, can be extended to arbitrary order and can be applied to other controllable driven systems with a residual coupling to a bath.

\subsection*{Decoherence in a rapidly driven nonlinear system}

The starting point of the calculation is the driven system-bath Hamiltonian
\begin{equation}
\label{eq:Htot}
\begin{aligned}
\hat{H}_{\mathrm{tot}}(t) = \hat{H}_{\mathrm{s}} + \hat{H}_{\mathrm{b}}+ \hat{H}_{\mathrm{sb}} +\hat{H}_{\mathrm{d}}(t). 
\end{aligned}
\end{equation}
The system is a weakly nonlinear oscillator whose Hamiltonian is given by $\hat{H}_{\mathrm{s}}/\hbar = \omega_{o} \hat{a}^{\dagger} \hat{a}+\sum_{n} \frac{g_{n}}{n}\left(\hat{a}+\hat{a}^{\dagger}\right)^{n}.$
Here, $\hat a$ is the bosonic annihilation operator. The parameters $\omega_o$ and $g_n\ll\omega_o$ are the bare oscillator frequency and the $n$-th rank nonlinearity coefficients of the oscillator.  We specialize our calculation to the case of the Josephson cosine potential as a source of oscillator nonlinearity and thus take the nonlinear coefficient $g_n$ of the Hamiltonian expansion to be of order $\varphi^{n-2}_{\mathrm{zps}}$ \cite{Venkatraman2021}, where $\varphi_{\mathrm{zps}}$ is the zero point spread of the phase across the Josephson junction $\hat \varphi = \varphi_\text{zps}(\hat a + \hat a^\dagger)$. The system is driven by $\hat{H}_{\mathrm{d}}(t) = -i\hbar F(t)\left(\hat{a}-\hat{a}^{\dagger}\right)$, where $F(t)$ is the waveform of the drive. At this time, we limit our analysis to the modeling of experiments in which the time dependence of the Hamiltonian corresponds to a monochromatic drive $F(t) = \Omega_d \cos(\omega_d t)$. The environment is taken to be a bath of linear oscillators with Hamiltonian $\hat{H}_{\mathrm{b}} = \sum_j \hbar \omega_j \hat{b}_{j}^{\dagger}\hat{b}_{j}$, which couples to the system by $\hat{H}_{\mathrm{sb}} = - \left(\hat{a}-\hat{a}^{\dagger}\right) \sum_j  h_{j} \left(\hat{b}_{j}-\hat{b}_{j}^{\dagger}\right)$. In these expressions $\hat{b}_{j}$ is the annihilation operator of a bath mode at frequency $\omega_j$.  

Motivated by the squeezed Kerr oscillator \cite{Wielinga1993, Cochrane1999, puri2017, puri2019, grimm2020, frattini2022,Chamberland2022, roberts2020}  and quantum information processing with cat-qubits \cite{puri2017,puri2019,grimm2020,leghtas2015, lescanne2020,Putterman2022,Chamberland2022,Gautier2022,Kwon2022}, we look now for the static effective description of $\hat{H}_\text{tot}$ under the condition $\omega_d\approx 2 \omega_o$. The construction of this effective description involves successive unitary transformations followed by averaging out the fast oscillation terms in the new frame. First, following \cite{Venkatraman2021}, we rewrite $\hat H_\text{tot}$ in a new frame comprising of (i) a displaced frame relative to the linear resonance of the oscillator to the drive so that \\ $\hat a \rightarrow \hat a + \frac{i\Omega_d}{2(\omega_d -\omega_o)}e^{-i\omega_dt}- \frac{i\Omega_d}{2(\omega_d +\omega_o)}e^{i\omega_dt}$, 
(ii) a rotating frame of $\hat a$ mode at $\omega_d/2$ so that $\hat a \rightarrow \hat a e^{-i\omega_dt/2}$, and then (iii) a rotating frame of each $\hat b_j$ mode at frequency $\omega_j$ so that  $\hat b_j \rightarrow \hat b_j e^{-i\omega_j t}$. The Hamiltonian now reads
\begin{subequations}\label{eq:Htot-dr}
\begin{align}
&\hat{H}_\text{tot} = \hat{H}_\mathrm{s}(t) + \hat{H}_{\mathrm{sb}}(t)\\
\begin{split}
&\hat{{H}}_{\mathrm{s}}(t)/\hbar=  \delta \hat{a}^{\dagger} \hat{a}+\sum_{n} \frac{g_{n}}{n}\Big(\hat{a} e^{-i \omega_d t/2 }+\hat{a}^{\dagger} e^{i \omega_d t/2}\\
&\qquad\qquad\qquad\qquad\qquad\quad+\Pi e^{-i \omega_{d} t}+\Pi^{*} e^{i \omega_{d} t}\Big)^{n} 
\end{split}\\
\label{eq:Hse}
\begin{split}
&\hat{{H}}_{\mathrm{sb}}(t)=i\left(\hat{a} e^{-i \omega_{d} t/2}-\hat{a}^{\dagger} e^{i \omega_{d} t / 2}\right)\hat B(t),
\end{split}
\end{align}
\end{subequations}
where $\delta=\omega_{o}-\omega_{d} / 2$, $\Pi \approx 4i\Omega_d / 3\omega_d$, and $\hat{B}(t) = \sum_j i h_{\omega_j}\left(\hat{b}_{j} e^{-i \omega_j t}-\hat{b}_{j}^{\dagger} e^{i \omega_j t}\right)$. Averaging out the fast oscillation arising in $\hat{{H}}_{\mathrm{s}}(t)$, one finds the system Hamiltonian and its coupling to the environment under the RWA (order $\varphi_\text{zps}^0$). We further replace the sum $\sum_j$ over the bath modes with an integral introducing a density of modes $\lambda_\omega$ such that $\lambda_\omega d\omega$ gives the number of oscillators with frequencies in the interval from $\omega$ to $\omega + d\omega$. Tracing out the environment at this point under the usual Born-Markov approximation in a thermal bath provides the ordinary Lindbladian \cite{Gautier2022,Putterman2022,Chamberland2022}, which involves the usual dissipators corresponding to single photon loss $\mathcal{D}[\hat{a}]$ and gain $\mathcal{D}[\hat{a}^{\dagger}]$ \cite{carmichael1999,breuer2002}, where $\mathcal D[\hat O]\bullet := \hat O\bullet\hat O^\dagger -(\hat O^\dagger \hat O\bullet + \bullet\hat O^\dagger \hat O)/2$. The effect of the bath under the Markov approximation is equivalent to a stochastic force coupled to the system by $i\hat f(t)(\hat a-\hat a^\dagger)$ with spectral density $S_{f\!f}[\omega] =2\pi \lambda_{\omega}|h_{\omega}|^2\bar n_\omega$,  $S_{f\!f}[-\omega] =2\pi \lambda_{\omega}|h_{\omega}|^2(1+\bar n_\omega)$, where $\bar n_\omega$ is the average photon number of the mode $\hat b_\omega$ at frequency $\omega>0$ \cite{clerk2010}.

The key to obtaining our main result is to take into account terms beyond the RWA in the system-bath coupling and get an averaged description of $\hat H_{\mathrm{tot}}$. We follow our generalization of the Schrieffer-Wolff transformation procedure \cite{Venkatraman2021} to construct a near-identity canonical transformation generated by $\hat{S}(t) = \mathcal O (\varphi_\text{zps})$ so that the transformed Hamiltonian is time-independent to order $\varphi_\text{zps}^k$ for some arbitrarily large $k$ of interest. 
Under $\hat S$, $\hat{H}_{\mathrm{tot}} (t) \rightarrow \hat{{\mathcal H}}_{\mathrm{eff}}$, which is given as
\begin{equation}
\begin{aligned}
\label{eq:Htot}
\hat{\mathcal{H}}_{\mathrm{eff}}&\equiv e^{\hat{S} / i \hbar} \hat{H}_{\mathrm{tot}}(t) e^{-\hat{S} / i \hbar}-i \hbar e^{\hat{S} / i \hbar} \partial_{t} e^{-\hat{S} / i \hbar}\\
&=\hat{\mathcal{H}}_{\mathrm{s}}+\hat{\mathcal{H}}_{\mathrm{sb}},
\end{aligned}
\end{equation}
where, by construction \cite{Venkatraman2021}, $\hat{{\mathcal H}}_{\mathrm{eff}}$ is the static effective approximation of $\hat{H}_{\mathrm{tot}} (t)$, and the computation of $\hat S(t)$ is detailed in \cref{supp:H-eff}. The first summand in \cref{eq:Htot} reads
\begin{equation}
\begin{aligned}
\label{eq:Hs2}
\hat{\mathcal{H}}_{\mathrm{s}}/\hbar = \Delta \hat{a}^\dagger \hat{a}-K\hat{a}^{\dagger 2}\hat{a}^2 +   \epsilon_2 (\hat{a}^{\dagger 2}+\hat{a}^2) + \mathcal{O}(\varphi_{\mathrm{zps}}^3),
\end{aligned}
\end{equation}
where $\Delta = \delta + 6g_4|\Pi|^2 - 18g_3^2|\Pi|^2/\omega_d + 2K$ is the Stark- and Lamb-shifted detuning, $K = -3g_4/2 +  20g_3^2/3\omega_d$ is the Kerr coefficient, and $\epsilon_2 = g_3\Pi$ is the squeezing amplitude. 

\section*{Effective Lindbladian at order $\varphi_\text{zps}^1$, i.e. first order beyond the RWA in the coupling to the environment}

The canonical transformation generated by $\hat S(t)$ can be viewed as describing the system in an accelerated frame. In this frame, the system effectively experiences the static Hamiltonian \cref{eq:Hs2}; meanwhile, the system-bath coupling develops nonlinear components. Keeping terms to order $\varphi_{\mathrm{zps}}^1$, the perturbation parameter in the expansion of $\hat{S}(t)$, the system-environment coupling reads
\begin{equation}
\begin{aligned}
\label{eq:Hse2}
\hat{\mathcal{H}}^{(1)}_{\mathrm{sb}}\approx{}&i\left(\hat{a} e^{-i \omega_{d} t / 2}-\hat{a}^{\dagger} e^{i \omega_{d} t / 2}\right) \hat{B}(t) \\
& +i\Big(-\frac{3 \epsilon_{2}}{\omega_{d}} \hat{a}^{\dagger} e^{i 3 \omega_{d} t / 2}-\frac{8 g_{3}}{3 \omega_{d}} \hat{a}^{\dagger 2} e^{i 2 \omega_{d} t / 2}\\
&-\frac{2 \epsilon_{2}}{\omega_{d}} \hat{a} e^{i \omega_{d} t / 2}+\frac{2 \epsilon_{2}}{\omega_{d}} \hat{a}^{\dagger} e^{-i \omega_{d} t / 2}\\
&+\frac{8 g_{3}}{\omega_{d}} \hat{a}^{2} e^{-i 2 \omega_{d} t / 2}+\frac{3 \epsilon_{2}}{\omega_{d}} \hat{a} e^{-i 3 \omega_{d} t / 2}\Big) \hat{B}(t),
\end{aligned}
\end{equation}
where the first line, at order $\varphi_\text{zps}^0$, is identical to the coupling term \cref{eq:Hse}.

Following a standard Lindbladian derivation \cite{carmichael1999,carmichael2009,breuer2002}, but now with the renormalized system-bath Hamiltonian, we obtain the effective Lindblad master equation for the system up to order $\varphi^1_\text{zps}$ as
\begin{align}
\begin{split}\label{eq:drs4}
    \partial_t\hat{\rho}_s =& \frac{1}{i\hbar}[\hat{\mathcal H}_s,\hat{\rho}_s]+ \kappa_{\omega_{d} / 2} \bar{n}_{\omega_{d} / 2}\mathcal{D}[\hat{a}^{\dagger} +\frac{2\epsilon_2}{\omega_d}\hat{a}]\hat{\rho}_{\mathrm{s}} \\
    &+\kappa_{\omega_{d} / 2}\left(1+\bar{n}_{\omega_{d} / 2}\right)\mathcal{D}[\hat{a} +\frac{2\epsilon_2}{\omega_d}\hat{a}^{\dagger}]\hat{\rho}_{\mathrm{s}} \\
    &+\kappa_{\omega_{d}} \bar{n}_{\omega_{d}}\left(\frac{8 g_{3}}{3 \omega_{d}}\right)^{2} \mathcal{D}\left[\hat{a}^{\dagger 2}\right] \hat{\rho}_{\mathrm{s}}\\
    &+ \kappa_{\omega_{d}}\left(1+\bar{n}_{\omega_{d}}\right)\left(\frac{8 g_{3}}{3 \omega_{d}}\right)^{2} \mathcal{D}\left[\hat{a}^{2}\right] \hat{\rho}_{\mathrm{s}} \\
    &+ \kappa_{3 \omega_{d} / 2} \bar{n}_{3 \omega_{d} / 2}\left(\frac{3 \epsilon_{2}}{\omega_{d}}\right)^{2} \mathcal{D}\left[\hat{a}^{\dagger}\right] \hat{\rho}_{\mathrm{s}}\\
    &+ \kappa_{3 \omega_{d} / 2}\left(1+\bar{n}_{3 \omega_{d} / 2}\right)\left(\frac{3 \epsilon_{2}}{\omega_{d}}\right)^{2} \mathcal{D}[\hat{a}] \hat{\rho}_{\mathrm{s}}.
\end{split}
\end{align}
Here, $\kappa_{\omega} = 2\pi \lambda_{\omega}|h_{\omega}|^2/\hbar^2 = (S_{f\!f}[-\omega] - S_{f\!f}[\omega])/\hbar^2$ is the system-bath coupling rate at frequency $\omega$.

As our first observation, we note that one can expand the dissipator $\mathcal{D}[\hat{a} +\frac{2\epsilon_2}{\omega_d}\hat{a}^{\dagger}]$ in \cref{eq:drs4} to find a \textit{heating} term that remains finite even at \emph{zero temperature}: $ \kappa_{\omega_{d} / 2}\Big(\frac{2\epsilon_2}{\omega_d}\Big)^2\mathcal{D}[\hat{a}^{\dagger}]$.  Its physical origin is a drive photon at frequency $\omega_d$ being converted to an oscillator excitation and an environment excitation, both at $\omega_d/2$. The associated effective Unruh-like temperature grows with the squeezing amplitude. 

The dominant correction for the situation that interests us, however, is the parity-preserving two-photon heating term $\mathcal{D}[\hat{a}^{\dagger2}]$. Its physical origin is in the thermal fluctuation at frequency $\omega_d$ driving incoherently the parametric process engineered to generate squeezing \cite{grimm2020,frattini2022}.

\begin{figure}[t!]
    \centering
    \includegraphics{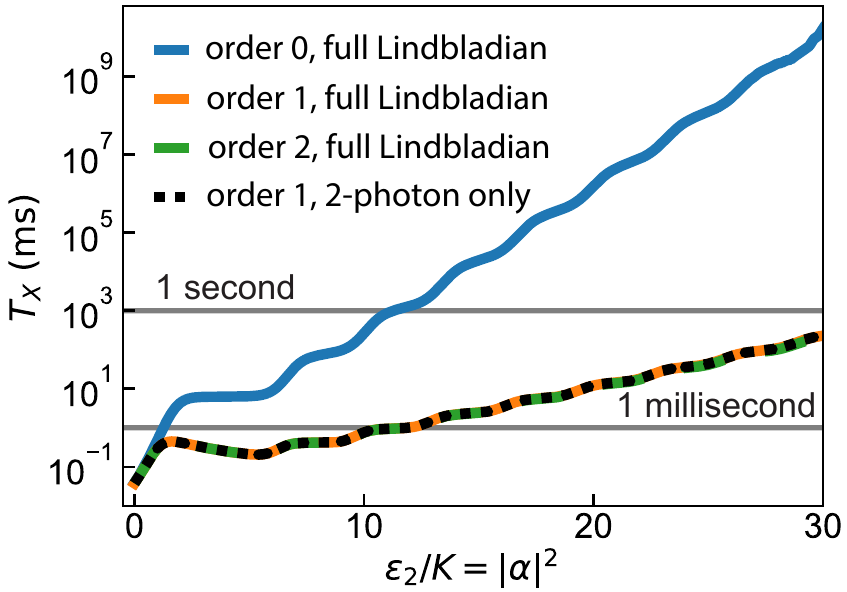}
    \caption{\textbf{$T_X$ vs ${|\alpha|^2}$ for different orders of perturbation theory and realistic parameters}. Theoretical predictions for the coherent state lifetime $T_X$ of the Kerr-cat qubit to  order $\varphi^0_{\mathrm{zps}}$ in the coupling to the environment, corresponding to the ordinary Lindbladian treatment (containing only single photon loss and gain at $\omega = \omega_d/2$, blue), to order $\varphi^1_{\mathrm{zps}}$ in the coupling to the environment (orange), and to order $\varphi^2_{\mathrm{zps}}$ (green). Also shown is the effect of keeping only the two-photon terms at order $\varphi^1_{\mathrm{zps}}$ (see \cref{eq:simdrs4}, black).}
    \label{fig:convergence}
\end{figure}

With the Lindbladian at order $\varphi^1_\text{zps}$, we can compute the decoherence time $T_X$ of the ground coherent states (a.k.a Glauber states) of the system $\big|\alpha = \pm\sqrt{\epsilon_2/K}\big\rangle\approx |\pm X\big\rangle$, where $X$ stands for a Bloch sphere axis \cite{grimm2020,frattini2022}. This quantity is the smallest non-zero real part of the Lindbladian eigenspectrum \cite{albert2014,frattini2022}. In \cref{fig:convergence}, we plot this quantity as a function of the squeezing amplitude. For our simulation parameters, we take $\kappa_{\omega_d}=5$~\textmu s$^{-1}$ and temperature $T_{\omega_d}=350$~mK, which are reasonable values for a drive port considering standard couplings and the noise temperature of the electronics controlling the microwave signals in quantum circuit experiments. In addition, we choose $\kappa_{\omega_d/2}=\kappa_{3\omega_d/2}=0.05$~\textmu s$^{-1}$ and temperature $T_{\omega_d/2}=T_{3\omega_d/2}=50$~mK, which are also based on experimental conditions of interest for this note. The nonlinear coefficients are taken to be $g_3/6\pi=20$~MHz, $g_4/ 8\pi=  280$~kHz, and consequently $K/2\pi = 320$~kHz, which are standard values for the SNAIL transmon \cite{frattini2018} used in the experiments \cite{frattini2022}. The drive frequency is $\omega_d/2\pi = 12$~GHz and the renormalized detuning in \cref{eq:Hs2} is taken to be $\Delta = 0.$

In \cref{fig:convergence}, we show the Lindbladian prediction for the ordinary dissipators (order $\varphi^0_{\mathrm{zps}}$, blue) and that for dissipators to order $\varphi^1_{\mathrm{zps}}$ (orange). The two predictions disagree by several orders of magnitude, and thus the former, being incomplete, is unfit to describe state-of-the-art experiments \cite{frattini2022}. The prediction to order $\varphi^2_{\mathrm{zps}}$ (green), which we discuss in detail next, adds negligible corrections and shows the convergence of the method for the chosen parameter values. We note that the ratio of the prefactors of two-photon heating at order $\varphi_\text{zps}^1$ and single photon heating at order $\varphi_\text{zps}^0$ is $17~\mathrm{Hz}/12~\mathrm{Hz}\sim 1$. Yet the two-photon process becomes dominant for $\epsilon_2/K>2$ because its strength scales as $\langle(\hat a^{\dagger}\hat a) ^2\rangle\sim |\alpha|^4$ while that of the single photon process scales as $\langle\hat a^{\dagger}\hat a\rangle\sim |\alpha|^2$ . We also plot the Lindbladian prediction (black), computed from 
\begin{align}
    \begin{split}\label{eq:simdrs4}
    \partial_t\hat{\rho}_s = &\frac{1}{i\hbar}[\hat{\mathcal H}_s,\hat{\rho}_s]+\kappa_{\omega_{d} / 2} \bar{n}_{\omega_{d} / 2}\mathcal{D}[\hat{a}^{\dagger}]\hat{\rho}_{\mathrm{s}} \\
&+\kappa_{\omega_{d} / 2}\left(1+\bar{n}_{\omega_{d} / 2}\right)\mathcal{D}[\hat{a} ]\hat{\rho}_{\mathrm{s}} \\
&+\kappa_{\omega_{d}} \bar{n}_{\omega_{d}}\left(\frac{8 g_{3}}{3 \omega_{d}}\right)^{2} \mathcal{D}\left[\hat{a}^{\dagger 2}\right] \hat{\rho}_{\mathrm{s}}\\
&+ \kappa_{\omega_{d}}\left(1+\bar{n}_{\omega_{d}}\right)\left(\frac{8 g_{3}}{3 \omega_{d}}\right)^{2} \mathcal{D}\left[\hat{a}^{2}\right] \hat{\rho}_{\mathrm{s}},
\end{split}
\end{align}
which only adds to the linear dissipators the term $\propto~\mathcal{D}[\hat{a}^{\dagger2}]$ (and its conjugate). Its close similarity with the full Lindbladian prediction confirms that two-photon heating constitutes the dominant corrections to the ordinary Lindbladian. Note, though, that this decoherence process has only a marginal effect on the lifetime of large Schr\"odinger cat states ($\propto~|\alpha \rangle \pm  |-\alpha \rangle$), since it conserves the parity of the state. Despite the failure of the ordinary Lindbladian to predict the lifetime of the coherent states, the lifetime of the Schr\"odinger cat states measured in \cite{frattini2022} is still accounted for by the ordinary linear dissipation because of its inherent fragility to single photon loss events.

\section*{Effective Lindbladian at order $\varphi_\text{zps}^2$, i.e. second order beyond the RWA in the coupling to the environment}

Similarly to the computation done at order $\varphi_\text{zps}^1$, we also compute $\hat S(t)$ generating the unitary transformation \cref{eq:Htot} to order $\varphi_\text{zps}^2$, as well as the effective Lindbladian to this order. The full expression is given in \cref{eq:order-2-full} in the appendix. The correction to this order may become relevant depending on the choice of parameters in the model, as we now discuss.

The second order Lindbladian samples the noise spectrum at $5\omega_d/2$, $2\omega_d$ and near zero frequency in addition to those sampled at the lower orders. For the noise spectrum at these frequencies, we chose $\kappa_{5\omega_d/2}=\kappa_{2\omega_d/2}=50$~ms$^{-1}$ and $T_{5\omega_d/2}=T_{2\omega_d}=50$~mK. For zero frequency, we take $\kappa_0 = 0$. These assignations were used also for the calculation to order $\varphi_\text{zps}^2$ in \cref{fig:convergence}. We remark that the assignation for $\kappa_0$ is an important assumption, justified for the decoherence model proposed here. For a thermal bath of linear oscillators, the number of photons diverges near DC as $\bar n_\text{th}\sim k_BT/\hbar\omega$ while the density of modes (and thus $\kappa_{\omega}$) goes to zero as a polynomial in $\omega$ ($\propto\omega^2$ for a resistance coupled to the circuit by a capacitance). Thus, the noise spectral density at near-DC frequency goes to zero as $\omega\rightarrow0$ in this model. However, for other noise models better suited to describe the low-frequency band including, for example quasi-particle loss and inductive loss \cite{pop2014,masluk2013,smith2020}, the noise near DC could become dominant and \cref{eq:Hse} should also be extended to capture the corresponding coupling terms. 

\begin{figure}[t!]
    \centering
    \includegraphics{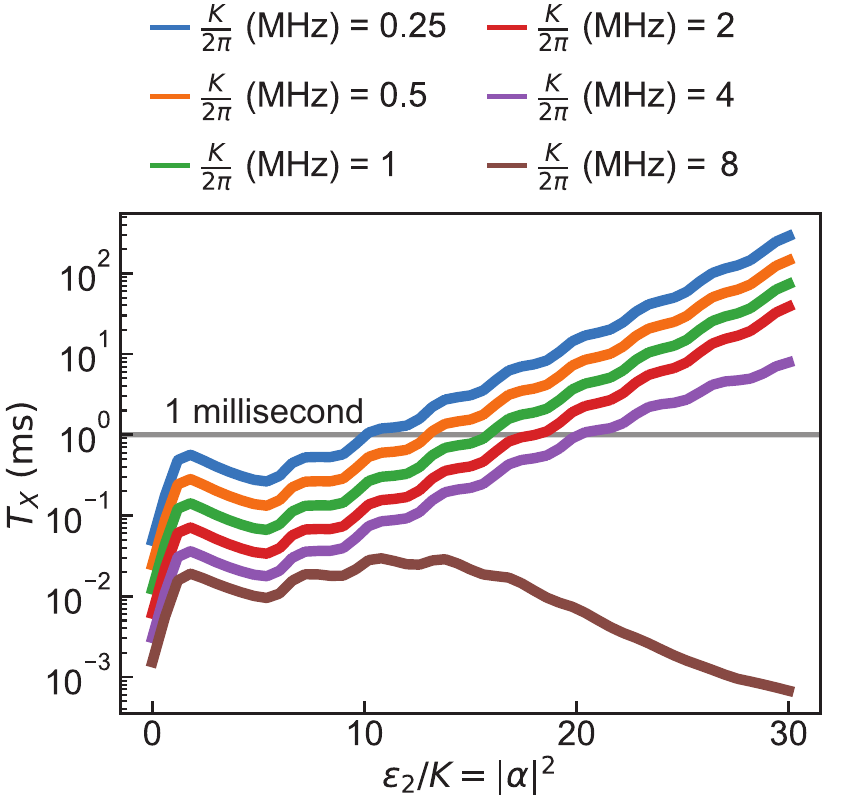}
    \caption{\textbf{$T_X$ vs ${|\alpha|^2}$ for different Kerr nonlinearities}. Theoretical predictions for the coherent state lifetime $T_X$ of the Kerr-cat qubit to order $\varphi_\text{zps}^2$ as a function of its mean photon number for several values of the Kerr constant $K$. The third rank nonlinearity is kept constant for all curves at $g_3/6\pi = 20~\mathrm{MHz}$. The bath parameters are identical to \cref{fig:convergence}. The brown curve corresponds to the prediction for parameters close to those in \cite{grimm2020}, whereas the blue curve corresponds to the prediction for parameters close to those in \cite{frattini2022}.}
    \label{fig:gns}
\end{figure}

In \cref{fig:gns}, we show the effect of increasing the Kerr nonlinearity in the lifetime prediction at order $\varphi_\text{zps}^2$ while keeping the rank-three nonlinearity constant to $g_3/6\pi = 20~\mathrm{MHz}$ as in \cite{grimm2020, frattini2022}. The Kerr coefficient is varied by varying $g_4$ \cite{frattini2018}. The dominant dissipator appearing at this order is $\mathcal{D}[\hat{a}^{\dagger}\hat{a}]$, which on the coherent states acts as a single photon gain enhanced by a factor $|\alpha|^2$. The magnitude of this dissipator scales as $|K|^4|\Pi|^2\langle (\hat{a}^{\dagger}\hat{a})^2\rangle\propto |K|^4 |\alpha|^8$ (see \cref{eq:order-2-full}) and its prefactor ranges between $10^{-6}$ and $2$ times that of the dissipator $\mathcal{D}[\hat{a}^{\dagger 2}]$ at order $\varphi_\text{zps}^1$ when $K/2\pi$ is varied from 0.25 MHz to 8 MHz for a coherent state with $|\alpha|^2 = 20$. Consequently,  this term becomes dominant for $K/2\pi>2$ MHz and for sufficiently large coherent state amplitudes. This is in qualitative agreement with the fact that the device in \cite{grimm2020}, characterized by $K/2\pi= 6.7$~MHz, has a $T_X$ lifetime considerably lower than the one achieved in \cite{frattini2022} where the device was operated at $K/2\pi= 320$~kHz.\footnote{Note that in order to achieve a given large Kerr ($\gg\kappa_{\omega_d/2}$), and thus fast gates in the Kerr-cat qubit, one should reduce as much as possible the decoherence induced by $g_3$ and $g_4$. One sees from the analytical expression in Eq.~(\ref{eq:order-2-full}) that the prefactors of some dissipators can be minimized or even cancelled, at constant Kerr, by the proper choice of the oscillator's nonlinearities.}

The main point of the exploration presented in this note is to showcase that an in-depth theoretical understanding of the dissipative processes at various orders is necessary for the experimental activity on parametric processes, like amplification, driven qubits, and quantum gates.

\subsection*{Comparison between theoretical predictions and experimental results}

\begin{figure}[t!]
    \centering
    \includegraphics{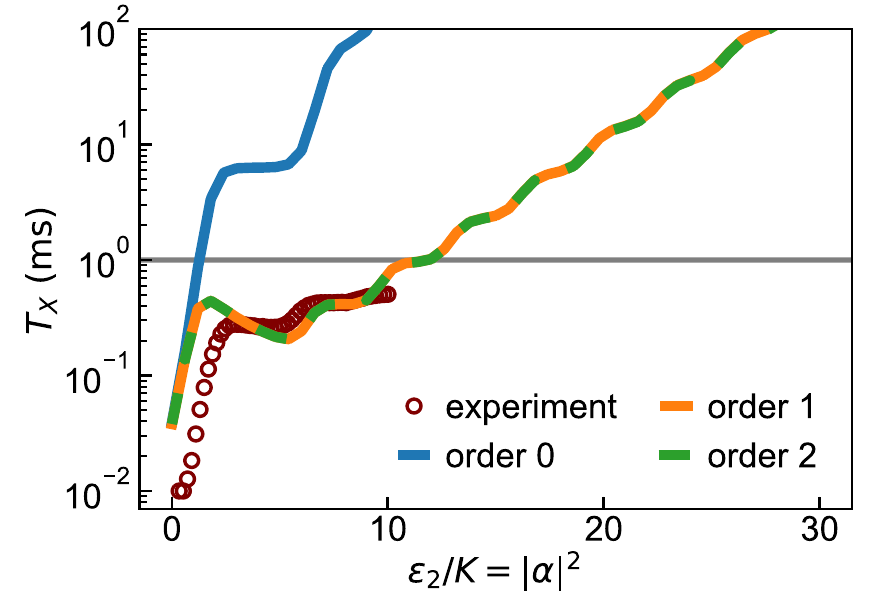}
    \caption{\textbf{Experimental data and model}. The experimental points correspond to the data in \cite{frattini2022}. The bath parameters in the model are identical to \cref{fig:convergence}.}
    \label{fig:data}
\end{figure}

We have found that an ordinary Lindbladian treatment is incomplete by several orders of magnitude when the beyond-RWA terms are examined for the Kerr-cat qubit. Consequently, to account for experimental observations, higher orders in the Lindbladian need to be considered. With the analytical expression presented here, we are able to account for the order of magnitude of the observations presented in \cite{frattini2022}, which are reproduced as maroon dots in \cref{fig:data}. Note that for $|\alpha|^2<2$, where there is a discrepancy between the experimental results and the predictions presented here,  the data has been explained in \cite{frattini2022}, by the inclusion of non-Markovian low-frequency noise which is not included here. The results presented in this note emphasise the need for further experiments that will in turn lead to detailed modeling of possible noise sources affecting particularly driven qubits.

\subsection*{Acknowledgements}
We acknowledge Alec Eickbusch, Daniel K. Weiss, Qile Su, Shruti Puri, and Steven M. Girvin for useful comments.

\appendix

\section{Mitigating lifetime reduction by adding two-photon cooling}
\label{subsec2phcool}
\begin{figure}[t!]
    \centering
    \includegraphics{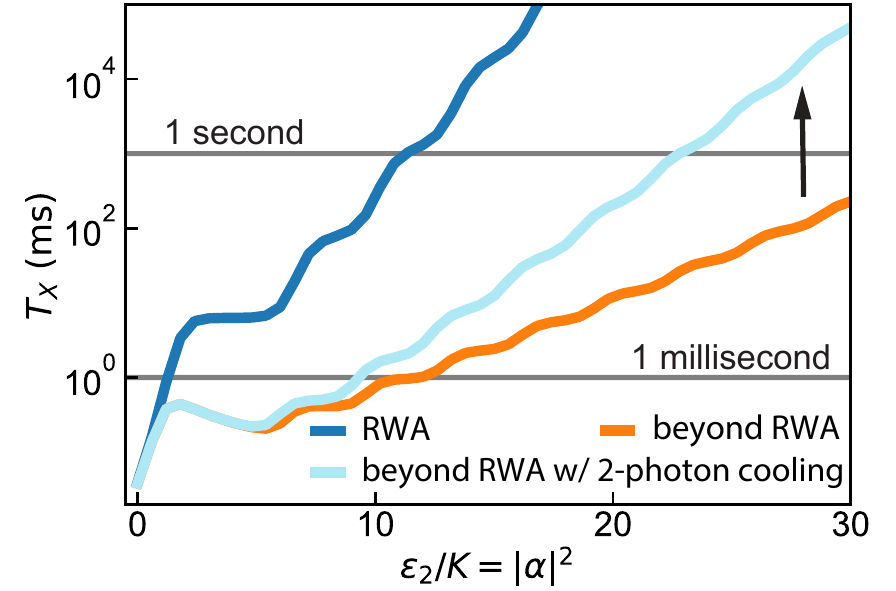}
    \caption{\textbf{Effect of two-photon cooling}. Adding an artificial two-photon dissipation term with a relatively small prefactor to the Kerr-cat system largely compensates the effect of the higher-order dissipators computed here. System and bath parameters have been chosen to be identical to those in \cref{fig:convergence}.}
    \label{fig:2ph}
\end{figure}

We identify that the dominant decoherence mechanisms are two-photon heating  $\mathcal{D}[\hat{a}^{\dagger 2}]$ and dephasing $\mathcal{D}[\hat{a}^{\dagger }\hat a]\propto |\alpha|^2\mathcal{D}[\hat a ^{\dagger}]$. To counteract this, we include, in our computation, a small amount of engineered two-photon dissipation \cite{mirrahimi2015,leghtas2015,puri2019,lescanne2020,doucet2020} ($\kappa_{\mathrm{2ph}}= 0.003$~\textmu s$^{-1}$). We show the outcome of the calculation in \cref{fig:2ph} for system and bath parameters as in \cref{fig:convergence,fig:data}. Experimentally, this should be easily achievable since much larger two-photon cooling rates have been demonstrated \cite{touzard2019,lescanne2020}, albeit in absence of a Kerr nonlinearity. Note, however, that a correct understanding is likely to require a higher-order analysis of engineered dissipation \cite{Gautier2022,Putterman2022} like the one presented here. 
It is likely that the combination of Hamiltonian stabilization and reservoir engineering will provide the agility and fast universal gates for cat-qubits and high coherent state lifetimes \cite{puri2019}.

\section{Static effective Hamiltonian}\label{supp:H-eff}

Here we follow \cite{Venkatraman2021} to compute $\hat S$ that generates the sought-after canonical transformation. First we expand \cref{eq:Htot} as 
\begin{subequations}
\begin{align}
    \hat{\mathcal{H}}_{\mathrm{eff}}&\equiv e^{\hat{S} / i \hbar} \hat{H}_{\mathrm{tot}}(t) e^{-\hat{S} / i \hbar}-i \hbar e^{\hat{S} / i \hbar} \partial_{t} e^{-\hat{S} / i \hbar}\\
    \begin{split}\label{eq:supp-bch}
     & =  \hat{{H}}_{\mathrm{s}}+ \frac{1}{i\hbar}[\hat S, \hat{{H}}_{\mathrm{s}}]+\frac{1}{2!(i\hbar)^2}[\hat S,[\hat S, \hat{{H}}_{\text{s}}]] + \cdots \\
     & \quad + \hat{{H}}_{\mathrm{sb}}+ \frac{1}{i\hbar}[\hat S, \hat{{H}}_{\mathrm{sb}}]+\frac{1}{2!(i\hbar)^2}[\hat S,[\hat S, \hat{{H}}_{\text{sb}}]] + \cdots \\
     & \quad + \partial_t\hat S  + \frac{1}{2! i\hbar} [\hat S, \partial_t {\hat S}] + \cdots
     \end{split}\\\label{eq:supp-eff-Hs-Hsb}
     & = \hat{\mathcal{H}}_{\mathrm{s}}  + \hat{\mathcal{H}}_{\mathrm{sb}},
\end{align}
\end{subequations}
where in \cref{eq:supp-bch} we have plugged in $\hat{{H}}_{\mathrm{tot}} = \hat{{H}}_{{s}}+ \hat{{H}}_{\mathrm{sb}}$ as defined in \cref{eq:Htot-dr} and employed the  Baker-Campbell-Hausdorff formula; in \cref{eq:supp-eff-Hs-Hsb} $\hat{\mathcal{H}}_{\mathrm{sb}}$ corresponds to the second line of \cref{eq:supp-bch} and $\hat{\mathcal{H}}_{\mathrm{s}}$ consists of the rest of \cref{eq:supp-bch}, which contains no bath modes.

Our goal is to perturbatively find $\hat S$ so that in the correponding frame $\hat{\mathcal H}_\text{s}$ is time-independent to some desired order of $\varphi_\text{zps}$. We therefore write $\hat{{H}}_{\mathrm{s}}$ and $\hat S$ each as a series 
\begin{align}
    \hat{{H}}_{\mathrm{s}} = \sum_{k>0} \hat{{H}}_{\mathrm{s}}^{(k)}, \quad \hat S = \sum_{k>0} \hat S^{(k)}
\end{align}
where $\hat{{H}}_{\mathrm{s}}^{(k)}$ and $\hat S^{(k)}$ are the order $\varphi_\text{zps}^k$ components in the corresponding series. 

Demanding $\hat{\mathcal H}_\text{s}$ to be time-independent at order $\varphi_\text{zps}^1$ \cite{Venkatraman2021}, we obtain the first order generator of the static effective transformation as 
\begin{align}
\frac{\hat S^{(1)}}{\hbar}=\;&- \int dt\; \text{osc}\left(\hat{H}_\text{s}\right)\\
\begin{split}
=\;&\frac{2}{5}  i \frac{g_{3}}{\omega_{d}} a^{\dagger} \Pi^{* 2} e^{i 5 \omega_{d} t / 2}+\frac{1}{2} i \frac{g_{3}}{\omega_{d}} a^{\dagger 2} \Pi^{*} e^{i 4 \omega_{d} t / 2} \\
&+\left(\frac{2}{3} i \frac{g_{3}}{\omega_{d}} a \Pi^{* 2}+\frac{2}{9} i \frac{g_{3}}{\omega_{d}} a^{\dagger 3}\right) e^{i 3 \omega_{d} t / 2}\\
&+2 i \frac{g_{3}}{\omega_{d}} a^{\dagger} a \Pi^{*} e^{i 2 \omega_{d} t / 2} \\
&+\Big(4 i \frac{g_{3}}{\omega_{d}}|\Pi|^{2} a^{\dagger}+2 i \frac{g_{3}}{\omega_{d}} a^{\dagger 2} a+2 i \frac{g_{3}}{\omega_{d}} a^{\dagger}\Big) e^{i \omega_{d} t / 2} \\
&+\text {h.c.}\raisetag{8pt}
\end{split}
\end{align}
where $ \text{osc}(f)= f-\int^T_0dt\;f$ extracts the oscillating part of $f$ with $T$ being its periodicity.

At this order, the transformed system-bath coupling is
\begin{align}
    \hat{\mathcal{H}}_{\mathrm{sb}}^{(1)} = \frac{[\hat S^{(1)}, \hat{{H}}_{\mathrm{sb}}]}{i\hbar},
\end{align}
where $\hat{{H}}_{\mathrm{sb}}$ is taken to be of order $\varphi_\text{zps}^0$. Carrying out the calculation explicitly, one then obtains $\hat{\mathcal{H}}_{\mathrm{sb}}^{(1)}$ in \cref{eq:Hse2}.

At order $\varphi_\text{zps}^2$, the generator of the canonical transformation is accordingly given by 
\begin{equation}
\begin{aligned}
 \frac{\hat S^{(2)}}{\hbar} \!=\!-\int_0^t dt \operatorname{osc}\!\Big(\hat H_\text{s}^{(2)} + \frac{[\hat S^{(1)},\hat{{H}}_{\mathrm{s}}^{(1)}]}{i\hbar} + \frac{[\hat S^{(1)}, \partial_t {\hat S^{(1)}}]}{2! i\hbar}\Big),
\end{aligned}
\end{equation}
the system-bath coupling is
\begin{align}
\begin{split}
\hat{{H}}_{\mathrm{sb}}^{(2)}&=\frac{1}{i \hbar}\left[S^{(2)},  \hat{{H}}_{\mathrm{sb}}(t)\right]\\
&+\frac{1}{2 !}\left(\frac{1}{i \hbar}\right)^{2}\left[S^{(1)},\left[S^{(1)}, \hat{{H}}_{\mathrm{sb}}(t)\right]\right],
\end{split}
\end{align}
and the full Lindbladian master equation up to this order is
\begin{widetext}
\begin{align}\label{eq:order-2-full}
\begin{split}
\partial_t{\hat{\rho}_{\mathrm{s}}} &=\frac{1}{i \hbar}\left[\hat{\mathcal{H}}^{(2)}_{\mathrm{s}}, \hat{\rho}_{\mathrm{s}}\right]+  \kappa_{0}\left(1+\bar{n}_{0}\right) \mathcal{D}\left[32 \frac{g_{3}^{2}}{\omega_{d}^{2}} a^{2} \Pi^{*}\right]+ \kappa_{0}\bar{n}_{0}\ \mathcal{D}\left[32 \frac{g_{3}^{2}}{\omega_{d}^{2}} a^{\dagger 2} \Pi\right] \\
&+  \kappa_{\omega_{d} / 2}\left(1+\bar{n}_{\omega_{d} / 2}\right)\left(\mathcal{D}\left[a+\frac{2g_3}{\omega_{d}} a^{\dagger}\Pi-\left(\frac{35}{2} \frac{g_{3}^{2}}{\omega_{d}^{2}}-6 \frac{g_{4}}{\omega_{d}}\right) a|\Pi|^{2}-\left({\frac{152}{9}} \frac{g_{3}^{2}}{\omega_{d}^{2}}-3 \frac{g_{4}}{\omega_{d}}\right) a^{\dagger} a^{2}-\left({\frac{152}{9}} \frac{g_{3}^{2}}{\omega_{d}^{2}}-3 \frac{g_{4}}{\omega_{d}}\right) a\right] \hat{\rho}_{\mathrm{s}}\right) \\
&+  \kappa_{\omega_{d} / 2} \bar{n}_{\omega_{d} / 2}\left(\mathcal{D}\left[a^{\dagger}+\frac{2 g_3}{\omega_{d}} a\Pi^*-\left(\frac{35}{2} \frac{g_{3}^{2}}{\omega_{d}^{2}}-6 \frac{g_{4}}{\omega_{d}}\right) a^{\dagger}|\Pi|^{2}-\left(\frac{152}{9} \frac{g_{3}^{2}}{\omega_{d}^{2}}-3 \frac{g_{4}}{\omega_{d}}\right) a^{\dagger 2} a-\left(\frac{152}{9} \frac{g_{3}^{2}}{\omega_{d}^{2}}-3 \frac{g_{4}}{\omega_{d}}\right) a^{\dagger}\right] \hat{\rho}_{\mathrm{s}}\right) \\
&+  \kappa_{\omega_{d}}\left(1+\bar{n}_{\omega_{d}}\right) \mathcal{D}\left[\frac{8 g_{3}}{3 \omega_{d}} a^{2}-\left(\frac{592}{9} \frac{g_{3}^{2}}{\omega_{d}^{2}}-16 \frac{g_{4}}{\omega_{d}}\right) a^{\dagger} a \Pi\right] \hat{\rho}_{\mathrm{s}} \\
&+  \kappa_{\omega_{d}} \bar{n}_{\omega_{d}} \mathcal{D}\left[\frac{8 g_{3}}{3 \omega_{d}} a^{\dagger 2}-\left(\frac{592}{9} \frac{g_{3}^{2}}{\omega_{d}^{2}}-16 \frac{g_{4}}{\omega_{d}}\right) a^{\dagger} a \Pi^{*}\right] \hat{\rho}_{\mathrm{s}} \\
&+  \kappa_{3 \omega_{d} / 2}\left(1+\bar{n}_{3 \omega_{d} / 2}\right) \mathcal{D}\left[\frac{3 g_3}{\omega_{d}} a\Pi-\left(\frac{51}{5} \frac{g_{3}^{2}}{\omega_{d}^{2}}-\frac{9}{2} \frac{g_{4}}{\omega_{d}}\right) a^{\dagger} \Pi^{2}+\left(4 \frac{g_{3}^{2}}{\omega_{d}^{2}}+\frac{3}{2} \frac{g_{4}}{\omega_{d}}\right) a^{3}\right] \hat{\rho}_{\mathrm{s}} \\
&+ \kappa_{3 \omega_{d} / 2} \bar{n}_{3 \omega_{d} / 2} \mathcal{D}\left[\frac{3 g_3}{\omega_{d}} a^{\dagger}\Pi^*-\left(\frac{51}{5} \frac{g_{3}^{2}}{\omega_{d}^{2}}-\frac{9}{2} \frac{g_{4}}{\omega_{d}}\right) a \Pi^{* 2}+\left(4 \frac{g_{3}^{2}}{\omega_{d}^{2}}+\frac{3}{2} \frac{g_{4}}{\omega_{d}}\right) a^{\dagger 3}\right] \hat{\rho}_{\mathrm{s}} \\
&+  \kappa_{2 \omega_{d}}\left(1+\bar{n}_{2 \omega_{d}}\right) \mathcal{D}\left[\left(\frac{224}{45} \frac{g_{3}^{2}}{\omega_{d}^{2}}+\frac{16}{5} \frac{g_{4}}{\omega_{d}}\right) a^{2}\right]\hat{\rho}_{\mathrm{s}}\\
&+  \kappa_{2 \omega_{d}}\bar{n}_{2 \omega_{d}} \mathcal{D}\left[\left(\frac{224}{45} \frac{g_{3}^{2}}{\omega_{d}^{2}}+\frac{16}{5} \frac{g_{4}}{\omega_{d}}\right) a^{\dagger 2}\right]\hat{\rho}_{\mathrm{s}}\\
&+  \kappa_{5 \omega_{d}/2}\left(1+\bar{n}_{5 \omega_{d}/2}\right) \mathcal{D}\left[\left(\frac{19}{9} \frac{g_{3}^{2}}{\omega_{d}^{2}}+\frac{5}{2} \frac{g_{4}}{\omega_{d}}\right) a^{2}\right]\hat{\rho}_{\mathrm{s}}\\
&+  \kappa_{5 \omega_{d}/2}\bar{n}_{5 \omega_{d}/2} \mathcal{D}\left[\left(\frac{19}{9} \frac{g_{3}^{2}}{\omega_{d}^{2}}+\frac{5}{2} \frac{g_{4}}{\omega_{d}}\right) a^{\dagger 2}\right]\hat{\rho}_{\mathrm{s}}.
\end{split}
\end{align}
\end{widetext}

One can also obtain the photon-number-dependence and Kerr-dependence of relevant terms above using the relationship $g_3\Pi = K|\alpha|^2$ and $K = -3g_4/2+20g_3^2/3\omega_d$. With this, one sees that the prefactor of $\mathcal D[\hat a^\dagger \hat a]$ at frequency $\omega_d$ is $\propto |K|^4$. Such strong dependence on $K$ explains the drastic drop in $T_X$ for $K/2\pi > 2$~MHz in \cref{fig:gns}, while for $K/2\pi < 2$~MHz, the effect of $\mathcal D[\hat a^\dagger \hat a]$, which is of order $\varphi_\text{zps}^2$, is much weaker than the effect of dissipators at lower orders and thus the change of the former is masked by that of the latter when $K$ varies in this regime. Note that by engineering the Hamiltonian nonlinearities $g_3$ and $g_4$, one may be able to mitigate the effect of these dissipators even for a system with large $K$.

\section{Future directions and refinement of the model}

When deriving the effective Lindbladian, we have made a few important assumptions. 

First, we note that we use the usual Born approximation which amounts to assuming $h_j \ll \omega_o \varphi_\text{zps}^2$.
This is, the Born approximation induces an error $\mathcal O(h_j)$ which needs to remain much smaller than the perturbative corrections computed, which are of order $\varphi_\text{zps}^2$ in this work. Under the same assumption, we demand the transformed system-bath Hamiltonian $\hat{\mathcal H}_\text{eff}=\hat{\mathcal H}_\text{s}+\hat{\mathcal H}_\text{sb}$ to be static to order $\varphi_\text{zps}^2$. But since $\hat{\mathcal H}_\text{sb} = \mathcal O(h_j)$, this amounts to demand that only $\hat{\mathcal H}_\text{s}$ be static, which provides an important but nonessential simplification.

We also remark that, in the standard Born-Markov approximation \cite{carmichael1999}, one treats the system-bath coupling term in the interaction picture, i.e. $\exp({-\mathcal{\hat H}_\text{s}/i\hbar})\mathcal{\hat H}_\text{sb}\exp({\mathcal{\hat H}_\text{s}/i\hbar})$, instead of $\mathcal{\hat H}_\text{sb}$ as we did in this work. The omission of this frame transformation is valid under the assumption that the bath is white in the neighbourhood of any given frequency $\omega_j$ with a width of a few $K$'s wide covering the relevant portion of the spectrum of $\hat{\mathcal H}_\text{s}$. This assumption holds generally for $\omega_j\gg K$ \cite{carmichael1999,carmichael2009}, but should be dealt with delicately for the near-DC noise, which may be treated numerically. Specifically, one can numerically compute the DC system-bath coupling in the interaction picture defined by $\hat{\mathcal H}_s$ in \cref{eq:Hs2}. This will transform the DC system-bath coupling to a sum of near-DC terms. One can subsequently trace out the bath under the Born-Markov approximation and obtain the effective Lindbladian.

\bibliography{bib}
\end{document}